\documentstyle[11pt,pasp,twoside,epsf]{article}
\def\edcomment#1{\iffalse\marginpar{\raggedright\sl#1\/}\else\relax\fi}
\reversemarginpar

\begin{document}
\title{X-ray source population of the Galactic center region obtained with ASCA}
 \author{Masaaki Sakano}
\affil{NASDA/SURP, Tsukuba, Ibaraki 305-8505, Japan,
sakano@oasis.tksc.nasda.go.jp}
 \author{Katsuji Koyama, Hiroshi Murakami}
\affil{Kyoto Univ., Sakyo, Kyoto, 606-8502, Japan}
 \author{Yoshitomo Maeda}
\affil{Pennsylvania State Univ., University Park, PA 16802-6305, USA}
 \author{Shigeo Yamauchi}
\affil{Iwate Univ., Morioka, Iwate, 020-8550, Japan}
 \author{The ASCA Galactic plane/center survey team}

\begin{abstract}
   From the {\it ASCA} X-ray point-source list in the Galactic center
 $5\times 5$ degree$^2$ region,
 we found the clear correlation between the position of the sources
 and the absorption.  This fact implies that
 the major part of the absorption is due to the cold interstellar matter (ISM)
 in the line of sight. Using the correlation, we estimate the distribution
 of the cold ISM.
  We also found that the ratio of high mass binaries to low mass binaries is
 significantly smaller than that in the whole Galaxy or SMC, which implies
 that the past starburst activity in the Galactic center region was rather quiet.
\end{abstract}

\section{The {\it ASCA} Galactic Center Survey}

   Due to the crowdedness of and heavy absorption to the Galactic center
 (hereafter, GC) region, the information, including the distances,
 on the X-ray sources there has been limited.
  If the spectra of the sources are examined,
 the obtained column density can be a good indicator of the distance. 

   For the {\it ASCA} GC survey data,
 we have made point-source search, spectral fitting for the detected
 sources, and identification, combining past references (Sakano et al. 2001).
  We hence could determine their spectral parameters, including the column density,
 for the bright sources (Sakano et al. 2001).

   We here investigate what determines the column density to
 bright X-ray sources, and then estimate the distribution of cold interstellar matter
 in the GC region.  We study and discuss
 the statistical property of the X-ray sources in the GC region.
  We assume the distance to the GC to be 8.5 kpc.

\section{Column Densities}

\vspace*{-1.3cm}
\begin{figure}[h]
\begin{center}
\begin{minipage}[t]{0.48\textwidth}
\vspace*{-4.0cm}

 We found a good correlation of
 the column densities of bright sources to
 the absolute galactic latitudes (Fig.~1),
  whereas they are not constrained from
 the source category.
  These suggest that the large part of the source column
 densities are probably the cold ISM distributed
 along the plane and in the GC region.
 In addition, the little scatter implies that most of the sources are
 distributed in the special site, definitely near the GC.
  This relation reflects the cold ISM distribution in and in front of
 the GC region.  From Fig.~1,
\end{minipage}~~
\begin{minipage}[t]{0.48\textwidth}
\mbox{\plotone{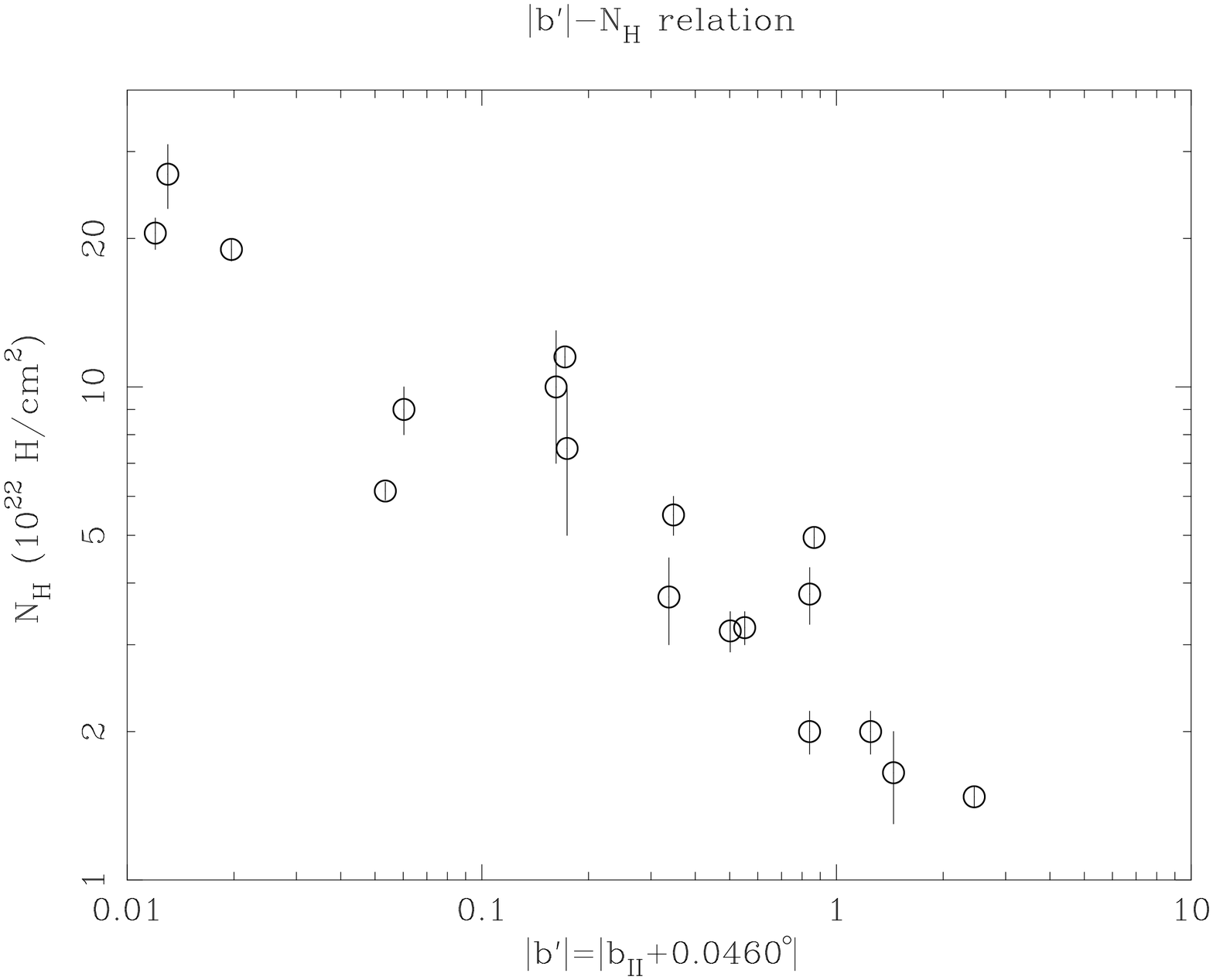}}
\caption{Source column density vs. angular distance from the plane (deg).}
\end{minipage}
\end{center}
\end{figure}
\vspace*{-0.83cm}

\noindent
 we thus
 estimate the scale length of the cold ISM density in the GC region
 for the perpendicular direction from the
 plane to be 20~pc, whereas the total cold
 ISM mass is calculated to be $6\times 10^7$ M$_{\odot}$,
 assuming the solar abundance and a simple radially-symmetrical shape for
 the cold ISM distribution.


\section{X-ray source population}

\begin{table}[tbp]
 \caption[]{X-ray source population \label{tbl:popu}}
  \begin{tabular}{cccccl}
 \hline
 & LMXB & HMXB & H/(H+L)$^{\dag}$ & mass & References\\
 \hline
GC & 23 & 4 &  15\% & 1 & this work\\
\hline
Galaxy & 79 & 34 & 30--40\% & 100 & van Paradijs 1995\\
SMC &  0 & 25 (+8) & 100\% & 1 & Yokogawa et al. 2000\\
 \hline
 \end{tabular} 

\footnotesize
\noindent $^{\dag}$: ratio of HMXB number to the total source number.
\end{table}

   Our results strongly confirmed the population of late-type objects
 in the GC region (Table~1).  
  Note that we classified pulsars to high-mass X-ray binaries (HMXBs)
 and bursters to low-mass X-ray binaries (LMXBs)
 because the companion stars are not identified for most of the sources.
 Since the companions of HMXBs are considered to have been born $\sim 10^7$
 years ago, the number of HMXBs or the ratio of HMXBs to LMXBs is a good indicator
 for the past star formation activity in the observed region;
  our result suggests rather quiet star formation activity $\sim 10^7$
 years ago in the GC region.




\begin{references}
\reference van Paradijs 1995, In ``X-ray Binaries'' (Cambridge: UP)
\reference Sakano, M., Koyama, K., Murakami, H., Maeda, Y., \& Yamauchi, S. 2001, \apjs, submitted
\reference Yokogawa, J., Imanishi, K., Tsujimoto, M., Nishiuchi, M., Koyama, K., Nagase, F., \& Corbet, R. H. D. 2000, ApJS, 128, 491
\end{references}
\end{document}